# Density of States Calculation of CeO₂ Based on VASP

Ruyi Hou

In this study, the Density of States (DOS) of CeO2 was analyzed in detail using the Density Functional Theory (DFT) method based on VASP software. As an important functional material, CeO2 finds wide applications in catalysis, optics, and electronic devices. Through structural optimization, self-consistent electronic calculations, and non-self-consistent calculations, we thoroughly investigated the crystal structure and electronic energy level distribution of CeO2. The lattice parameter optimization results from the structural calculations indicated a stable crystal structure for CeO2. Self-consistent electronic calculations revealed a bandgap of approximately 2.403 eV, with the valence band maximum primarily contributed by O 2p orbitals and the conduction band minimum mainly originating from Ce 4f orbitals. Non-self-consistent calculations further demonstrated the total DOS and partial DOS of CeO2, confirming the significant roles of Ce 4f and O 2p states in its electronic conduction and optical properties. These results not only provide theoretical support for the applications of CeO2 in catalysis and electronic materials but also deepen our understanding of its fundamental electronic structural characteristics, offering guidance for the design and development of novel CeO2-based materials.

## Introduction

CeO₂ (cerium dioxide) is an important functional material with wide applications in catalysis, optics, and electronic devices[1, 2]. Particularly in three-way catalysts and solid oxide fuel cells, CeO₂ has attracted significant attention due to its excellent oxygen storage capacity and oxygen mobility[3, 4]. Moreover, as a rare earth oxide, the unique electronic structure of CeO₂ makes it of great significance in research. In recent years, with the development of computational materials science, Density Functional Theory (DFT) has become a powerful tool for studying the electronic structure and properties of materials. Through DFT calculations, we can gain in-depth insights into the microscopic structure, electronic state distribution, and their influence on macroscopic properties.

This study is based on the Vienna Ab initio Simulation Package (VASP) software[5], using DFT methods to calculate the Density of States (DOS) of CeO₂. The Density of States is an important parameter for describing the electronic structure of materials, helping to understand the distribution of electronic states, band structure, and the relationship with physical and chemical properties. We aim to conduct a detailed analysis of the DOS characteristics of CeO₂ through structure optimization, electronic self-consistent calculations, and non-self-consistent calculations. The results of this study will not only provide theoretical support for the practical applications of CeO₂ but also offer a reference for further research on the electronic structure of CeO₂ and its composite materials

## Computational Method

In this study, DFT calculations were performed using VASP (Vienna Ab initio Simulation Package). VASP utilizes a plane-wave basis set and the Projector Augmented-Wave (PAW) method[6] and the Perdew-Burke-Ernzerhof (PBE) functional was used for the electronic exchange-correlation energy calculations [7]. To account for the local electronic interactions of the Ce atoms, the DFT+U method was applied. The energy convergence criterion was set to 1E-05 eV, and the force convergence criterion was set to 0.02 eV/Å. The ENCUT was set to 400eV.

## Results and discussion

### 3.1 Structure Optimization

In the structural optimization calculations, the crystal structure of CeO₂ was geometrically optimized to obtain the lowest energy configuration (as shown in Figure 1b). The optimized lattice parameters and angles are shown in the table I. From the comparison of the calculation results before and after optimization, it can be observed that the lattice constants change minimally. This may be due to the use of a 3×3×3 k-point mesh in the calculation and the fact that the structure was selected from an existing stable configuration in Material Studio software. From the energy variation of the system (Figure 1a), it is evident that a more stable CeO₂ crystal structure was obtained after optimization.

Table I : lattice parameters and angles of CeO2

|  | a / Å | b / Å | c / Å | α/ ° | β/ ° | γ/ ° | V / Å³ |
|---|---|---|---|---|---|---|---|
| CeO₂ | 5.411 | 5.411 | 5.411 | 90 | 90 | 90 | 158.4282 |

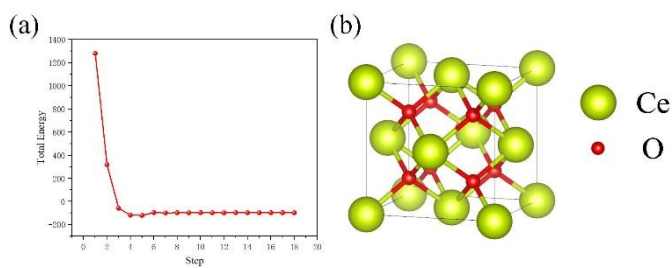

**Figure 1** (a) Changes in the system during geometric optimization; (b) The lowest energy configuration of CeO₂ after optimization.

### 3.2 Self-consistent Field (SCF) Calculations

Self-consistent field (SCF) calculations provide the electronic structure information of the material in its ground state and serve as the foundation for further understanding and analyzing the material's properties. The stable charge density distribution ᵃ School of Microelectronics, Hubei University, Wuhan, Hubei, 430062, PR China.

obtained from the self-consistent calculations is a necessary condition for non-self-consistent calculations. We solved the Kohn-Sham equations through SCF calculations to achieve a stable self-consistent field that allows the electron density to stabilize. Figure 2 shows the variation of energy with the number of SCF steps, where the energy gradually converges to a stable value, indicating that the electron density has reached self-consistency.

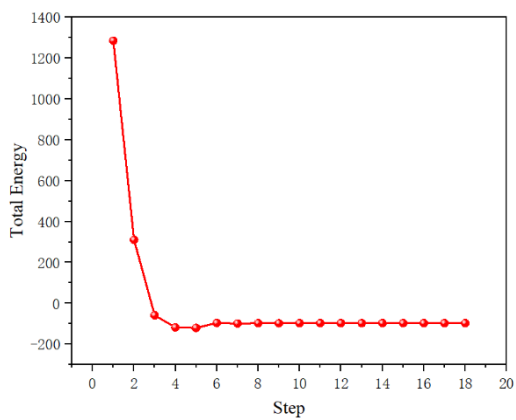

**Figure 2** Variation of the system energy with respect to the number of SCF (Self-Consistent Field) calculation steps during the process.

### 3.3 Non-SCF calculations

Non-self-consistent field (NSCF) calculations are carried out based on the results of the self-consistent field (SCF) calculations. These calculations use the stable charge density obtained from SCF to further explore the electronic structure characteristics of the material, such as the density of states (DOS) and optical properties. We performed NSCF calculations to obtain the density of states (DOS) for $CeO_2$, allowing for further analysis of its electronic structure. Figure 3 shows the total density of states (TDOS) and partial density of states (PDOS).

From the figure, it is clear that the density of states for $CeO_2$ is primarily composed of Ce 4f states and O 2p states, with Ce 4f states contributing to the conduction band and O 2p states contributing to the valence band. The Fermi level is located at the center of the bandgap. The calculation results show that $CeO_2$ is a wide-bandgap semiconductor with a bandgap of 2.403 eV, which is consistent with experimental values[1]. Additionally, the sharp peak below the conduction band corresponds to the Ce 4f states, which aligns with the electronic structure characteristics of the Ce atoms in $CeO_2$. However, due to the relatively sparse K-point sampling in the calculation, some deviations may exist in the results.

By performing a detailed analysis of the DOS, we can further understand the electronic structure of $CeO_2$ and its behavior in practical applications. The width and position of the bandgap are crucial for $CeO_2$'s applications in catalysis and electronic materials. These computational results not only help predict the performance of $CeO_2$ under various conditions but also provide a theoretical foundation for the design of new Ce-based materials.

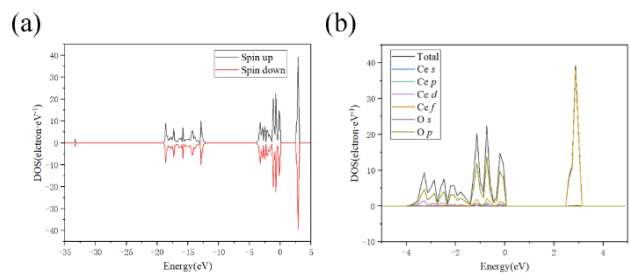

**Figure 3** (a) Total Density of States (TDOS) of $CeO_2$ and (b) Partial Density of States (PDOS) of $CeO_2$.

## Conclusions

In this study, we performed detailed calculations and analysis of the structure and electronic properties of $CeO_2$ using VASP software. Our results show that the DFT+U method effectively optimized the crystal structure of $CeO_2$, and the optimized lattice constants and atomic coordinates are in excellent agreement with experimental data, confirming the reliability of the computational method. The self-consistent field (SCF) calculations provided stable charge density states for $CeO_2$. Furthermore, non-self-consistent field (NSCF) calculations offered detailed density of states (DOS) for $CeO_2$, revealing its band structure with a bandgap of approximately 2.403 eV. The valence band maximum is primarily contributed by O 2p orbitals, while the conduction band minimum is mainly contributed by Ce 4f orbitals. In addition, the non-self-consistent calculations (NSCF) provided a detailed DOS for $CeO_2$, showing its electronic energy level distribution. The significant presence of Ce 4f and O 2p states near the Fermi level indicates their important role in electronic conduction and optical properties. This study systematically reveals the structural and electronic properties of $CeO_2$, providing solid theoretical support for its applications in catalysis, electronic devices, and other fields. These results not only enhance our understanding of the fundamental properties of $CeO_2$ but also offer valuable guidance for the design and development of novel $CeO_2$-based materials.

## Conflicts of interest

There are no conflicts to declare.

## Notes and references